\documentclass{aastex63}

\newcommand{\xgw}{x_{\rm GW}}
\newcommand{\xem}{x_{\rm EM}}
\newcommand{\thetac}{\vec{\theta}_c}

\newcommand{\thetav}{\vec{\theta}}
\newcommand{\thetaem}{\vec{\theta}_{\rm EM}}
\newcommand{\evp}{\langle v_p \rangle}
\newcommand{\pr}[1]{p\left( #1 \right)}
\newcommand{\dLs}{\tilde{d}_L}
\newcommand{\zs}{\tilde{z}}
\newcommand{\D}{\mathrm{d}}
\newcommand{\ZD}{}

\published{March 20, 2020}
\submitjournal{ApJL}

\shorttitle{Thunder and Lightning}
\shortauthors{Z.~Doctor}
\graphicspath{{./}{figures/}}
\usepackage{amsmath}

\begin{document}

\title{Thunder and Lightning: Using Neutron-Star Mergers as Simultaneous Standard Candles and Sirens to Measure Cosmological Parameters}

\author[0000-0002-2077-4914]{Zoheyr Doctor}
\affiliation{Department of Physics\\
University of Oregon\\
Eugene, OR 97403, USA}
\email{zoheyr.doctor@gmail.com}

\begin{abstract}

With the detection of gravitational wave GW170817 and its associated electromagnetic counterparts from a binary neutron star merger, the ``standard siren" method for Hubble-constant measurements is expected to play a role in the Hubble-constant tension in the next few years.  One intriguing proposal put forward in multiple studies is to use a neutron-star merger's optical counterpart, known as a kilonova, as a standard candle, since its absolute magnitude can in principle be calculated from simulations. In this work, I detail the statistical framework for performing joint standard candle-standard siren measurements using gravitational waves, electromagnetic follow-up data, and simulations of electromagnetic counterparts. I then perform an example analysis using GW170817 and its optical counterpart AT2017gfo to illustrate the method and the method's limitations. Crucially, the inferences using this method are only as robust as the electromagnetic counterpart models, so significant theoretical advances are needed before this method can be employed for precision cosmology. 

\end{abstract}


\section{Introduction} \label{sec:intro}

With tensions mounting in the cosmology community over the value of the Hubble constant \citep{Freedman2017,Riess2016,Planck2018}, new and independent measurements of cosmological parameters have gained significant interest. The most provocative of these new measurements employs gravitational waves (GWs) from compact-object mergers and their associated electromagnetic (EM) counterparts as ``standard sirens" to estimate $H_0$, the Hubble constant \citep{Schutz,HolzHughes2005,DelPozzo2014,H0170817,darksiren,CFH2018,Mortlock2018}. These standard-siren measurements differ from traditional distance-ladder measurements in that the luminosity distance to the source can be directly inferred from the GW signal without the need of empirical calibration of increasingly distant sources.   For GW170817, the first gravitational wave detected from a neutron-star merger and the first to have associated EM counterparts, \citet{H0170817} inferred $H_0=70.0_{-8.0}^{+12.0}$ (maximum a-posteriori with 68\% credible interval), which, while exciting, was alone not enough to significantly tip the scales on the Hubble-constant controversy. \ZD{The second reported GW of potential neutron-star binary origin, GW190425 \citep{GW190425}, did not have an identified host galaxy \citep[e.g.][]{Hosseinzadeh,GROWTH190425}, and therefore offered no additional $H_0$ constraints.} Nevertheless, combining the GW170817 estimate with future standard-siren analyses is expected to yield a competitive $H_0$ measurement \ZD{with $O(100)$ GWs with identified host galaxies \citep{CFH2018}.}  Furthermore, EM-counterpart morphology can be leveraged to infer the inclination angle of the binary inspiral and break the well-known distance-inclination degeneracy, leading to improved estimates of $H_0$ \citep{Guidorzi,H0170817,Dhawan2019}.  

Until recently, no one had considered leveraging the optical data from a neutron-star (NS) merger to directly infer the source luminosity distance, enabling a standard-candle measurement using an EM counterpart.  New studies in \citet{KRA2019} and \citet{CoughlinH02019} have explored this possibility and treated the prompt, thermal radiation from neutron-rich ejecta from the merger, known as a kilonova (KN), as a standard candle.  Using simulations of neutron-star merger ejecta, \citet{KRA2019} find a clear relationship between the slope and the peak brightness of the KN bolometric light curves under certain assumptions, suggesting that KNe could be ``standardized".  \citet{CoughlinH02019} find similar correlations between light-curve stretch and brightness in KN simulations and use fits to these correlations to infer $H_0$ with GW170817 as a standard candle.  Also, they combine the $H_0$ posteriors of the GW170817 standard candle and standard siren analyses to produce a joint $H_0$ fit.  While these analyses have crucially laid the ground work for joint GW-EM inference of $H_0$, neither has presented the full Bayesian approach for performing these inferences with an arbitrary EM counterpart.

In this {\it Letter}, I present a method for performing joint GW-EM $H_0$ inferences and enumerate some of the technical subtleties which can affect the measurements.  In this method, the luminosity distance to the source is simultaneously fit using the gravitational wave with the light curves and/or spectra of the EM counterparts, drawing an analogy with estimating the distance to a lighting strike using both the brightness of the lightning and the loudness of the thunder. (Though unlike lightning and thunder, light and GWs travel at the same speed). This basically amounts to including the EM-counterpart {\it likelihood} in the standard-siren calculation\footnote{\ZD{Note that the calculation is {\it not} a simple multiplication of EM and GW-based distance posteriors as claimed in \citet{CoughlinH02019}.}}.  By including this likelihood, the luminosity distance to the source (and hence the cosmological parameter inference) is further constrained by the observed EM-counterpart morphology.

The layout of this {\it Letter} is as follows: In \S\ref{sec:methods}, I present the Bayesian framework for performing ``thunder and lightning" inferences of $H_0$. In \S\ref{sec:example} I perform an example thunder-and-lightning $H_0$ measurement with GW170817 and its bolometric light curve.  One essential take-away from \S\ref{sec:example} is that systematic errors in the underlying EM-counterpart models can significantly affect the cosmological inferences.  It is not clear that currently available models are sufficiently systematics-free to be employed for precision cosmology, so I do not attempt to stack multiple simulated events in this work, but future work could explore this. Finally in \S\ref{sec:discussion} and \S\ref{sec:conclusion}, I conclude by offering some high-level discussion on the method presented herein and future prospects.  

\section{Statistical Framework} \label{sec:methods}

I develop a Bayesian framework for making thunder-and-lightning inferences of the Hubble constant (and other cosmological parameters) which incorporates both gravitational-wave and EM-follow-up data.  The framework makes use of the following ingredients:
\begin{itemize}
    \item Data:
    \begin{itemize}
        \item Gravitational-wave data $\xgw$: The strain time series in $N$ different detectors $\{h_i(t)\}_{i=1}^{N}$. 
        \item Electromagnetic-counterpart data $\xem$: Counts or flux measurements of the EM counterpart. This could be in multiple bands and/or with multiple instruments. 
        \item Host-galaxy recessional velocity $v_r$: Measured through spectroscopy of the host galaxy.
        \item Peculiar velocity field $\evp$ of the host galaxy group: A measurement of the peculiar velocity field in the vicinity of the host galaxy.  In \citet{H0170817}, this was computed through a weighted average of the peculiar radial velocities of galaxies nearby the host.
    \end{itemize}
    \item Models/Assumptions:
    \begin{itemize}
        \item Gravitational waveform model: A model for the gravitational-wave strain as a function of intrinsic source parameters (masses,spins, and equation of state of the compact binary) and extrinsic parameters (e.g.~luminosity distance, sky position, inclination). \ZD{Contemporary examples include the models of \citet{DietrichSelfSpin} and \citet{Lackey}.}
        \item EM-counterpart model: A forward model for $\xem$ that is a function of the compact binary parameters.  This could be a hierarchical model which predicts some electromagnetic counterpart parameters (e.g.~ejected mass or jet opening angle) based on the binary parameters, and then predicts $\xem$ via the EM counterpart parameters. \ZD{For example, \citet{CoughlinMMPE}, \citet{CoughlinEOS}, and \citet{EstimatingDynamical} provide state-of-the-art forward models of KNe.}
        \item Recessional velocity model: A model for the observed recessional velocity of the host given its peculiar velocity and a cosmology. \ZD{See \citet{H0170817} for an example.}
        \item Peculiar velocity field model: A model for the measured peculiar velocity field in the vicinity of the host galaxy given the host galaxy peculiar radial velocity. \ZD{See \citet{H0170817} for an example.}
    \end{itemize}
\end{itemize}

With these pieces in hand, one can compute the likelihood of the data set given the cosmological parameters $\thetac$ and the models. Following \citet{H0170817} I can write the likelihood as:
\begin{equation}
    \mathcal{L}(\xgw,\xem,v_r,\evp|\thetac) = \frac{\pr{\xgw,\xem,v_r,\evp|\thetac}}{\int_{\rm det} \pr{\xgw,\xem,v_r,\evp|\thetac} d\xgw d\xem d v_r d\evp }
\end{equation}
The denominator term depends only on $\thetac$ and appropriately renormalizes the likelihood to be the likelihood over {\it detectable} events \citep{H0170817,CFH2018}.  The integral is taken over data sets which meet some detection threshold and represents the fraction of all possible data sets, conditioned on $\thetac$, which would be included in this analysis.  

To see the dependence of the data on the cosmological parameters, I expand the numerator term as follows:
\begin{eqnarray}
    \label{eqn:expandL}
    \pr{\xgw,\xem,v_r,\evp|\thetac} &=& \int \pr{\xgw,\xem,v_r,\evp,\thetav,d_L,z,i,v_p|\thetac}\D\thetav \D d_L \D z \D i \D v_p \nonumber\\
     &=& \int \pr{\xgw,\xem,v_r,\evp|\thetac,\thetav,d_L,z,i,v_p} \pr{\theta,d_L,z,i|\thetac}\pr{v_p}\D\thetav \D d_L \D z \D i \D v_p \nonumber\\
     &=& \int \pr{\xgw|\thetac,\thetav,d_L,z,i}\pr{\xem|\thetac,\thetav,d_L,z,i}\pr{v_r,\evp|\thetac,\thetav,d_L,z,i,v_p} \nonumber\\ 
     && \quad \times \pr{\thetav,d_L,z,i|\thetac}\pr{v_p}\D \thetav \D d_L \D z \D i \D v_p \nonumber\\
     &=& \int \pr{\xgw|\thetav,d_L,z,i}\pr{\xem|\thetav,d_L,z,i}\pr{v_r,\evp|z,v_p} \nonumber\\ 
     && \quad \times \pr{\thetav}\pr{z|d_L,\thetac}\pr{d_L}\pr{i}\pr{v_p}\D\thetav \D d_L \D z \D i \D v_p  
\end{eqnarray}
$d_L$, $z$, and $i$, and $v_p$ are the source's luminosity distance, redshift, inclination angle, and host-galaxy peculiar velocity respectively, and $\thetav$ is the array of remaining parameters that describe the binary system, such as total mass, mass ratio, spins, and equation of state. Going from the first to second line of Equation \ref{eqn:expandL}, I apply Bayes' Rule.  In the third line, I split individual likelihoods under the assumption that the GW data, EM data, and measured velocities are independent. Then in the final line I separate the priors on $\thetac$, $d_L$, and $i$ by assuming they are independent and remove the explicit $\thetac$ dependence of $\xem$, $\xgw$, $v_r$, and $\evp$ since those are fully specified by $d_L$ and $z$. Note that I assume that $v_p$ negligibly affects the measured GW or EM counterpart data.  $z$ is fully specified by $d_L$ and cosmology, so $\pr{z|d_L,\thetac} = \delta(\tilde{z} - z)$ where $\tilde{z} = z(d_L,\thetac)$.  This leaves:
\begin{eqnarray}
    \pr{\xgw,\xem,v_r,\evp|\thetac} &=& 
    \int \pr{\xgw|\thetav,d_L,\zs,i}\pr{\xem|\thetav,d_L,\zs,i}\pr{v_r,\evp|\zs,v_p} \nonumber\\ 
    &&{}\quad \times \pr{\thetav}\pr{d_L}\pr{i}\pr{v_p}\D\thetav \D d_L \D i \D v_p
\end{eqnarray}
Now I turn to the second term in the integrand which is the likelihood of the electromagnetic follow-up data. In general, it is not straightforward to predict the EM data directly from the binary parameters, but a number of studies have attempted to do this via EM counterpart ansatz parameters $\thetaem$.  For example, a number of models parameterize kilonova spectra in terms of the mass, velocity, and composition of the material ejected in the neutron-star merger \citep[e.g.][]{Kasen2017,Bulla2019,CoughlinEOS,Metzger2019}. These parameters can in turn be predicted from the binary system parameters with the help of neutron-star merger simulations. We can therefore expand the EM likelihood as:
\begin{eqnarray}
    \pr{\xem|\thetav,\dLs,\zs,i} &=& \int \pr{\xem,\thetaem|\thetav,\dLs,\zs,i}\D \thetaem \nonumber \\
    &=& \int \pr{\xem|\thetaem,\dLs,\zs,i}\pr{\thetaem|\thetav}\D \thetaem 
\end{eqnarray}
The term $\pr{\thetaem|\thetav}$ enables us to encode our uncertainty about the EM counterpart parameters based on the binary parameters, which may be helpful in cases where there is significant error or uncertainty in the ansatz simulations. However, in principle, $\thetaem$ should be fully predicted by $\thetav$, in which case $\pr{\xem|\thetav,\dLs,\zs,i} = \pr{\xem|\thetaem(\thetav),\dLs,\zs,i}$. Wrapping this all together into a final posterior distribution on the cosmological parameters yields
\begin{eqnarray}
    \pr{\thetac|\xgw,\xem,v_r,\evp} = \frac{\pr{\thetac}}{\beta(\thetac)}\int && \pr{\xgw|\thetav,d_L,\zs,i}\pr{\xem|\thetaem,d_L,\zs,i}\pr{\thetaem|\thetav} \nonumber\\
    &&{} \quad \times \pr{v_r,\evp|\zs,v_p} \pr{\thetav}\pr{d_L}\pr{i} \pr{v_p} \D \thetaem \D \thetav \D d_L \D i \D v_p
\end{eqnarray}
where
\begin{eqnarray}
    \beta(\thetac) =  \int \int_{\substack{\xgw, \xem,\\ v_r, \evp \in {\rm det}}} && \pr{\xgw,\xem,v_r,\evp|\thetav,\thetaem,d_L,\zs,i,v_p}\pr{\thetaem|\thetav} \nonumber\\  &&\times \pr{\thetav}\pr{d_L}\pr{i}\pr{v_p} \D \xgw \D \xem \D v_r \D \evp \D \thetaem \D \thetav \D d_L \D i \D v_p
\end{eqnarray}

\subsection{Selection Effects}

The denominator term in the likelihood $\beta(\thetac)$ that accounts for selection effects has been discussed in other work on these types of measurements \citep{H0170817,CFH2018}, and it can vary in computational complexity depending on the situation and assumptions made. In cases where the EM and GW detection probabilities are unaffected by cosmology (e.g.~at low redshift), the denominator term can be ignored, as was done in \citet{H0170817}. In the general case though, it must be computed, \ZD{because the cosmological parameters affect the GW signal frequency band (detector frame), the sky-map area, and the EM counterpart brightness in the search filters, and hence affect the probability of detection}.

Before expanding $\beta(\thetac)$, I first make some simplifying assumptions. I assume that detection of $\xem$ enforces detection of $v_r$ and $\evp$ and hence combine the recessional velocity data into the variable $\xem$. This is a reasonable assumption because if an EM counterpart is pinpointed, its host galaxy redshift can be readily measured spectroscopically.  A second stipulation I make is that the detection of electromagnetic data is conditioned on detection of GW data.  While it is possible to do subthreshold searches for GW events based on serendipitous detection of KNe \citep{Doctor2017,ScolnicKN}, gamma ray bursts \citep{GRBsubthreshold,Harstad:2013mda}, or other counterparts, the vast majority of joint GW-EM detections will come from a GW detection triggering an EM search.  With these assumptions, $\beta(\thetac)$ is:
\begin{eqnarray}
        \beta(\thetac) &=&  \int \int_{\substack{\xgw, \xem\\ \in {\rm det}}}  \pr{\xgw,\xem|\thetav,\thetaem,d_L,\zs,i}\pr{\thetaem|\thetav} \pr{\thetav}\pr{d_L}\pr{i}\pr{v_p}\D \xgw \D \xem \D \thetaem \D \thetav \D d_L \D i  \nonumber \\
        &=& \int \int_{\substack {\xgw \\ \in {\rm det}}} p_{\rm det}^{\rm EM}(\xgw,\thetaem,d_L,\zs,i)p(\xgw|\thetav,d_L,\zs,i) \pr{\thetaem|\thetav} \pr{\thetav}\pr{d_L}\pr{i} \pr{v_p} \D \xgw \D \thetaem \D \thetav \D d_L \D i \nonumber \\
        &=& \int p_{\rm det}^{\rm EM,GW}(\thetaem,\thetav,d_L,\zs,i)p(\thetaem|\thetav) \pr{\thetav}\pr{d_L}\pr{i} \pr{v_p}\D \xgw \D \thetaem \D \thetav \D d_L \D i
\end{eqnarray}
where

\begin{eqnarray}
    p_{\rm det}^{\rm EM}(\xgw,\thetaem,d_L,\zs,i) &=& \int_{\xem \in {\rm det}} \pr{\xem|\xgw,\thetaem,d_L,\zs,i} \D \xem \label{eqn:pdetem}\\
    p_{\rm det}^{\rm EM,GW}(\thetaem,\thetav,d_L,\zs,i) &=& \int_{\xgw \in {\rm det}} p_{\rm det}^{\rm EM}(\xgw,\thetaem,d_L,\zs,i) p(\xgw|\thetav,d_L,\zs,i) \D\xgw \label{eqn:pdetemgw}
\end{eqnarray}
Equations \ref{eqn:pdetem} and \ref{eqn:pdetemgw} are the probability of detection of the EM counterpart given the GW data, source properties, and cosmology, and the probability of detection of both the EM counterpart {\it and} the GW given the source properties and cosmology, respectively. 

I will not attempt to calculate $\beta(\thetac)$ for arbitrary experimental configurations here, and instead I will simply comment on what considerations must go into computing it, if it is needed.  As noted in \citet{H0170817}, $\beta(\thetac)$ can be ignored in cases where the GW selection effects dominate and the GW BNS detection horizon distance is small. The selection effect on GWs is largely driven by the GW signal-to-noise ratio $\rho$, which will not change appreciably for the slight redshifting of the GW signal expected at such small distances (below a few hundred Mpc). Therefore, there is no selection of the data set which is conditioned on the cosmological parameters, and one can safely ignore $\beta$.  However, for larger GW horizon distances, substantial redshifting of the GW signals (which depends on $\thetac$) can change detection prospects, and $\beta(\thetac)$ must be explicitly calculated.  

If we now consider sources which are detectable at cosmological distances, we must estimate $\beta(\thetac)$. While $\beta(\thetac)$ could be directly computed by e.g.~Monte Carlo integration if the full model and data can be simulated, it is advantageous to make some simplifying assumptions. First, for GW detection, a signal-to-noise ratio threshold $\rho_*$ can be used as a proxy for signal selection threshold.  For EM counterparts, there is no hope in pinpointing the counterpart and getting its host galaxy redshift without actively pointing a telescope at the relevant sky position.  A number of complicated factors come into whether a telescope is pointed at the counterpart (e.g.~a team's allotted observing time, weather conditions, camera field of view, etc.), but roughly speaking, \ZD{GW events with small localization areas are more likely to be in an instrument's field of view than events with large localizations} \citep[e.g.][]{CoughlinOpticalImplications,CoughlinTeamwork}\footnote{\ZD{In some cases significant resources could be leveraged to cover large sky areas \citep[e.g.][]{GROWTH190425}, but this is unlikely to be the norm if observing resources remain constant and the detection rate of NS binaries increases due to increased interferometer sensitivity.}}. The sky localization area of a GW signal approximately scales with $\rho^{-2}$ for moderately large $\rho$ \citep{Fairhurst2009}, so we can model the selection function on the EM counterpart as depending on $\rho$ rather than the full $\xgw$, which lets us replace instances of $\xgw$ in Equation \ref{eqn:pdetemgw} with $\rho$ and integrate with respect to $\rho$ on interval $[\rho_*,\infty]$. 

Now even if one or more telescopes are pointed at the true sky location of the event in question, it is not guaranteed that the EM counterpart will be detected, as the counterpart could be either misidentified or too dim \citep[e.g.][]{Kyutoku}.  This will depend sensitively on the observing strategies and detection pipelines that EM-follow-up groups employ, which again may be difficult to model.  But for concreteness, let's consider a simple case: There is one EM-follow-up campaign which uses fixed exposure time and filters and a fixed detection pipeline which has detection threshold $\zeta_*$ on detection statistic $\zeta$, for which any observation with $\zeta\geq\zeta_*$ would be included in the analysis.  The probability of EM detection assuming the instrument has been pointed at the source will not depend on the GW data or $\rho$, but it will depend on both the distance and redshift (hence a cosmology dependence) to the source since the source must at least be of significant brightness {\it in the relevant filters}. With these assumptions and a model of $\zeta$ given the source parameters, Equation \ref{eqn:pdetemgw} can be rewritten as:
\begin{equation}
    p_{\rm det}^{\rm EM,GW}(\thetaem,\thetav,d_L,i,\zs) =  \int_{\zeta_*}^\infty \pr{\zeta|\thetaem,d_L,i,\zs,\in {\rm FOV}} \left[\int_{\rho_*}^\infty  \pr{\in {\rm FOV}|\rho} p(\rho|\thetav,d_L,i,\zs) \D\rho \right]\D\zeta \label{eqn:pdetemgw_simp}
\end{equation}
The first term in the outer integral of Equation \ref{eqn:pdetemgw_simp} is the probability of getting EM detection statistic $\zeta$ given that the source was in the field of view of the EM instrument. This term depends on the specifics of the EM pipeline and the filters and exposure times chosen. The next term $\pr{\in {\rm FOV}|\rho}$ (in the integral in brackets) is the probability that the source was in the instrument field of view given the GW signal-to-noise ratio.  This depends on how the EM instrument's field-of-view is pointed on the sky.  In principle, Equation \ref{eqn:pdetemgw_simp} is calculable, but careful modeling of each term is needed if \ZD{the detection probability of sources changes across the prior range of the cosmological parameters}. 

\section{Example}\label{sec:example}
In this section, I perform a toy thunder-and-lightning $H_0$ inference using GW170817 and the AT2017gfo measured bolometric optical/near-infrared light curve from integration of the X-shooter spectra \citep{Smartt2017,Pian2017}. Given the cornucopia of modeling uncertainties in KN light curves and spectra, I opt to use a simple model for NS merger mass ejection and the associated KNe.  The example herein, which is primarily for illustrative purposes, can be readily extended to multi-band light-curve or spectral fits of KNe with more complex KN models, or even to other EM counterparts. I emphasize that other models, data, and assumptions could be made with respect to the following example, and that the choices made here were established to (a) demonstrate execution of the method and (b) demonstrate how reasonable changes to the underlying model significantly affect the inferences.

\subsection{Data}
Rather than performing full GW inference on the GW170817 strain, I re-weight the low-spin posterior samples from GW170817 provided by the LIGO-Virgo Collaboration \citep{GW170817}\footnote{https://dcc.ligo.org/LIGO-P1800061/public}.  For the bolemetric light curves, I integrate the de-reddened, de-redshifted X-Shooter AT2017gfo optical/infrared spectra \citep{Smartt2017,Pian2017} \footnote{http://www.engrave-eso.org/AT2017gfo-Data-Release/}. Since the event is at such low redshift (even assuming a wide prior range on $H_0$), using the de-redshifted data here introduces negligible bias. In general though, such de-redshifted data cannot be used because it has already assumed a cosmology.

\subsection{Model}
Since I use the GW170817 posterior samples, the GW model and priors are already specified by the choices made in the LVC low-spin-prior analysis \citep{GW170817}, but other prior choices could be made if desired.  To model the kilonova, I make the following assumptions, which are chosen mostly for simplicity of this example:
\begin{itemize}
    \item The kilonova ejecta velocity and opacity are fixed to $v=0.25c$ and $\kappa=1$ cm$^2$/g, respectively.  The ejecta mass is calculated in two ways:  
    \begin{enumerate}
        \item Using Equation 25 of \citet{Radice2018} (which is modified to fit ejecta mass rather than disk mass):
        \begin{equation}
            \frac{M_{\rm ej}}{M_\odot}(\tilde{\Lambda}) = {\rm max}\left\{  10^{-4}, 0.0202 + 0.0341\times \tanh\left( \frac{\tilde{\Lambda} - 538.8}{439.4} \right) \right\} \label{eqn:mej}
        \end{equation}
        $\tilde{\Lambda}$ is the binary tidal deformability parameter \citep[e.g.][]{GW170817}
        \item Using Equations 1 and 2 of \citet{Dietrich2017} to estimate a dynamical ejecta component and a disk wind ejecta prescription from \citet{CoughlinMMPE} with a disk mass to ejecta mass conversion factor of 0.3.   
    \end{enumerate}
    Here, the parameters describing the kilonova $\thetaem$ are fully specified by the binary parameters $\thetav$, so $\pr{\thetaem|\thetav}=\delta(\kappa - 1 {\rm cm}^2{\rm g})\delta(v_{\rm ej} - 0.25c)\delta(M_{\rm ej} - M_{\rm ej}(\tilde{\Lambda}))$.
    \item The kilonova bolometric luminosity is calculated for a single ejecta component using Equations 1 and 2 of \cite{KRA2019} and the prescriptions chosen therein. Furthermore, I assume that the KN bolometric luminosity has no viewing angle dependence \footnote{Models with viewing-angle dependence could be used and indeed would help break distance-inclination degeneracy \citep{Bulla2019}}.

\end{itemize}

\subsection{Results}
\begin{figure}
    \centering
    \includegraphics{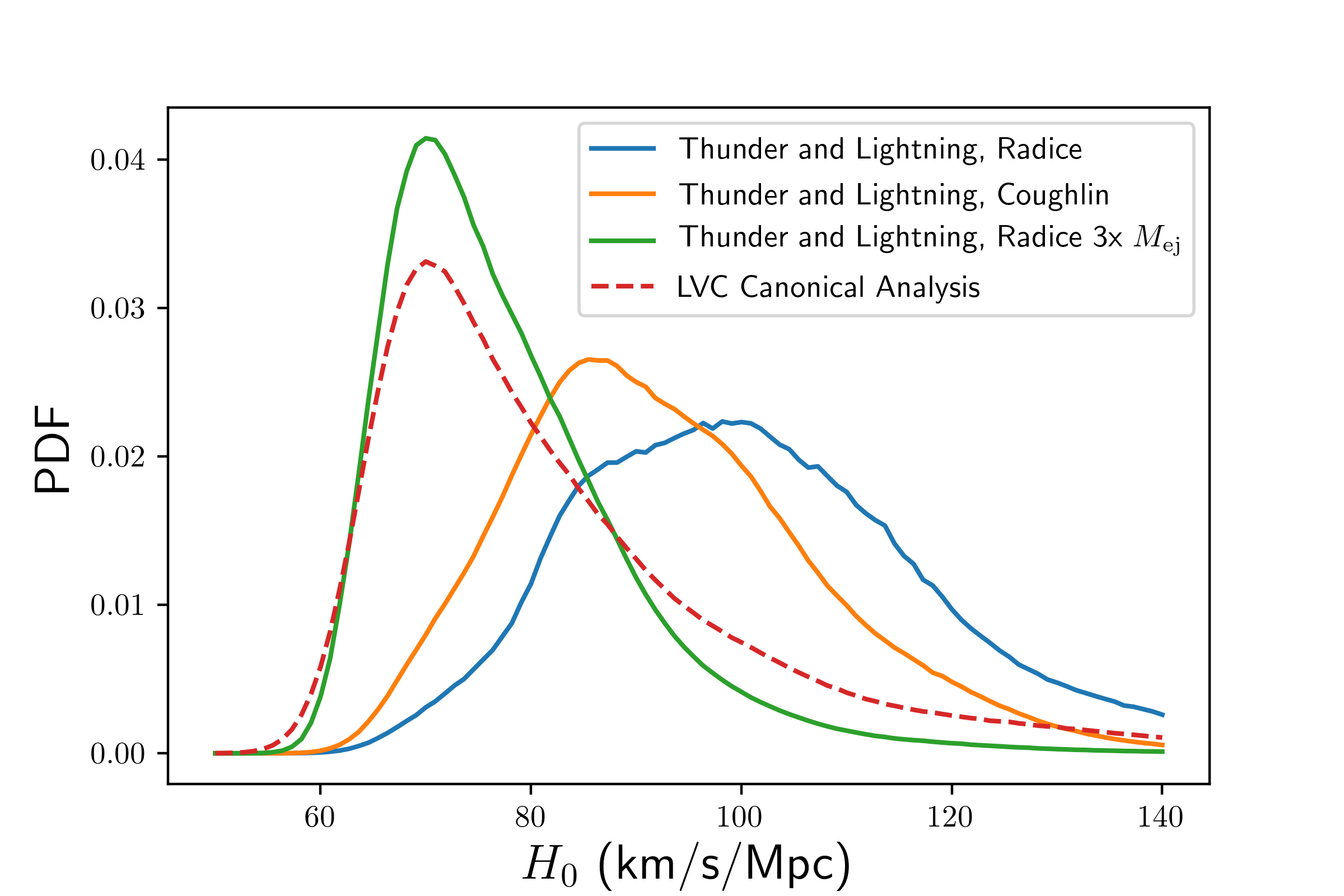}
    \caption{Posterior distributions on the Hubble constant $H_0$ for the LVC canonical analysis \citep[][but using the low-spin prior]{H0170817} and for thunder and lightning analyses. A thunder and lightning $H_0$ posterior is shown for the \citet{Radice2018} and \citet{CoughlinMMPE} ejecta mass calculations, and for an ad-hoc increase of their \citet{Radice2018} ejecta mass predictions of 3x.}
    \label{fig:H0post}
\end{figure}

To calculate the posterior distribution on $H_0$, I re-weight the GW170817 posterior samples by (a) the kilonova light curve likelihood, which I take to be $\chi^2$ in the bolometric magnitudes with constant $\sigma=1$ mag ``modeling uncertainty" (similarly to \citet{CoughlinH02019}), and (b) $\pr{v_r,\evp|\zs,v_p}$ through prior samples of $v_p$ and the prescriptions used in \citet{H0170817}. I use the same GW $H_0$ likelihood and velocity measurements as in \citet{H0170817} as well as the same priors ($p(d_L)\sim d_L^2$, $p(H_0)\sim 1/H_0$, flat in component masses). The thunder-and-lightning reweighting of the GW170817 posterior samples yields the posterior distributions on $H_0$ shown in Figure \ref{fig:H0post}.  The blue and orange curves show the thunder-and-lightning $H_0$ inference using The Radice and Coughlin kilonova ejecta prescriptions, respectively.  The green curve shows the results assuming three times the standard Radice ejecta mass. Note that these results incorporate the gravitational-wave $H_0$ inference (shown in the dashed red line) via the original LVC GW samples and therefore should not be multiplied again by the LVC canonical analysis posterior. 

\section{Discussion}\label{sec:discussion}
\begin{figure}
    \centering
    \includegraphics{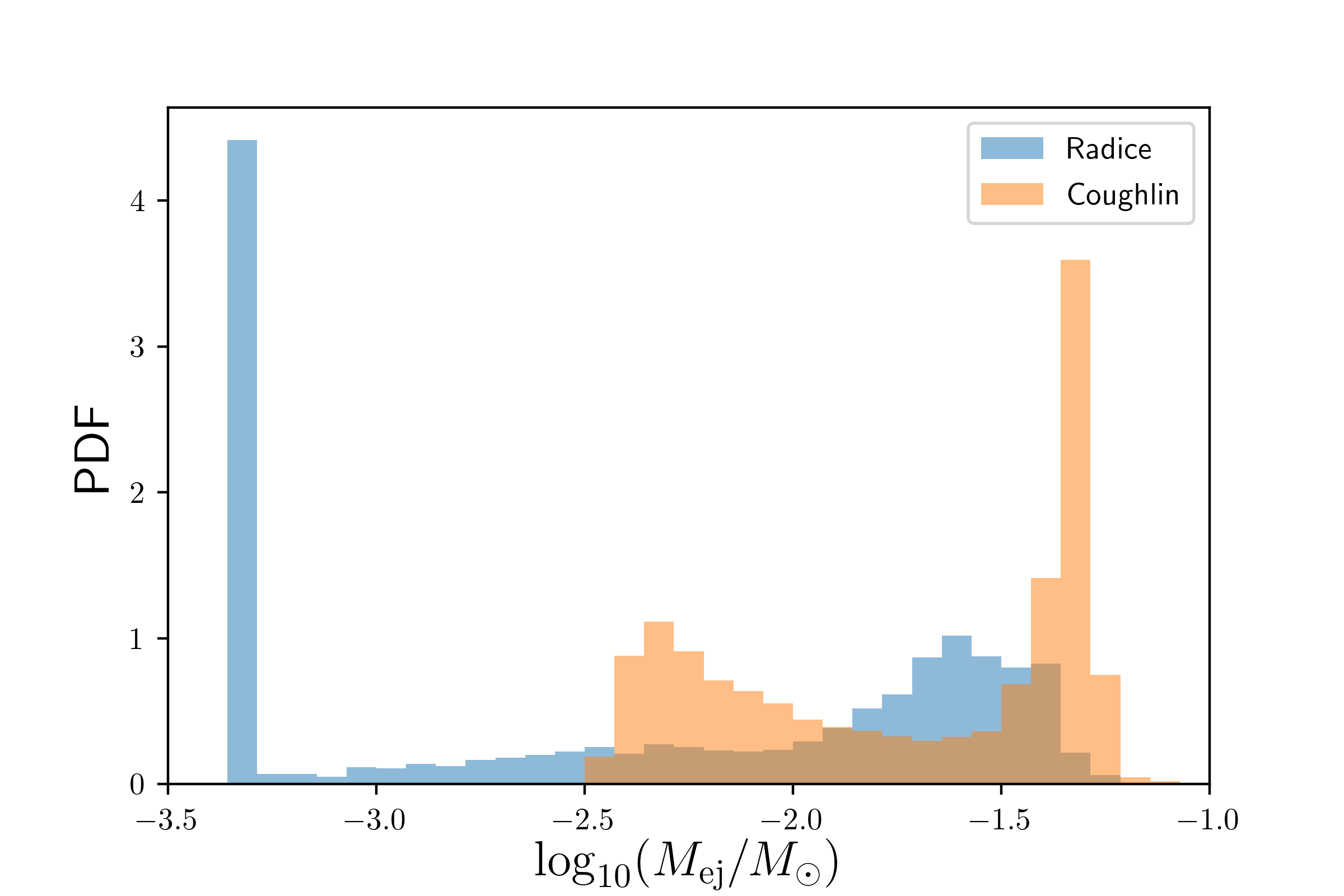}
    \caption{The ejecta masses calculated for the GW170817 posterior samples using prescriptions from \citet{Radice2018} and \citet{Dietrich2017}/\citet{CoughlinMMPE}.}
    \label{fig:mej}
\end{figure}

The comparison of $H_0$ posteriors shown in Figure \ref{fig:H0post} demonstrates the effects of different choices for the underlying EM model.  With the toy model presented herein coupled with the \citet{Radice2018} ejecta mass prescription, the $H_0$ posterior favors higher expansion rates than the pure standard-siren method.  This is due to the low ejecta masses predicted by Equation \ref{eqn:mej}, shown in Figure \ref{fig:mej}, which favor a small source luminosity distance and hence a higher $H_0$.  Using the ejecta masses from \citet{Dietrich2017} and \citet{CoughlinMMPE}, yields similar results, albeit less biased to large $H_0$ since the predicted ejecta masses are larger. If the ejecta masses from \citet{Radice2018} are arbitrarily increased by a factor of 3, the inference on $H_0$ becomes consistent with those in the existing literature.  It is also worth noting that these results would change if the high-spin analysis of the LVC were used instead.  The take-away from this toy example is that the model of the KN and $\pr{\thetaem|\thetav}$ are crucially important to the thunder and lightning $H_0$ inference.  I emphasize that the results shown here are not meant to be new constraints on the Hubble constant, but rather an illustration of the model dependence of thunder-and-lightning analyses.

\subsection{Systematics, Populations, and Self-Consistency}

\citet{CoughlinH02019} performed the first thunder-and-lightning $H_0$ measurement with optical light curves and state-of-the-art models and find general agreement between their results and existing studies.  However, as I have shown here, the underlying models used for the EM counterparts can greatly affect the Hubble constant measurement.  As such, careful accounting of modeling uncertainties must be done to recover unbiased $H_0$ estimates. For KNe for example, ejecta properties can vary from study to study \citep{Radice2018}, so $H_0$ estimates may have large systematic errors. Also, selection effects must be addressed when the GW detector horizon distances extend to distances at which cosmology can affect the joint GW-EM detection prospects. 

There are two other subtleties that must be addressed as well.  Firstly, if multiple joint GW-EM detections are brought to bear on $H_0$ in a combined thunder-and-lightning analysis, the compact-object-merger populations (e.g.~distribution of neutron-star masses) must be simultaneously fit with the cosmological parameters to account for degeneracy between the unknown population and the cosmology. This amounts to marginalizing over another set of variables $\vec{\lambda}$, which parameterize the NS mass and spin distributions. Without such a simultaneous fit, error in the assumed NS distributions will bias the inferred distances and hence the cosmology measurement.  

Second, the models that go into thunder-and-lightning analyses should not be conditioned or trained on existing analyses or data sets that assume an underlying cosmology. For example, many analyses of AT2017gfo used a known cosmology to infer properties of the KN ejecta.  If these inferred properties (e.g.~KN ejecta velocity profile) are assumed in future thunder and lightning analyses, the results will not be self-consistent due to existing cosmological assumptions creeping in. Therefore, the simulations and models used for thunder-and-lightning analyses should rely only on GR, particle/nuclear theory, and fits to experiments that do not involve cosmology. 

\section{Conclusion}\label{sec:conclusion}
In this {\it Letter}, I have expanded on the work of \citet{KRA2019} and \citet{CoughlinH02019} by showing the full Bayesian framework for thunder and lightning (joint GW-EM) inference of cosmological parameters.  Additionally, I have described subtleties of such inferences which were not discussed in these previous works.  In particular, thunder and lightning analyses must account for the following details:
\begin{itemize}
    \item Selection effects on the GW and EM data sets can potentially bias cosmological parameter measurements.  These selection effects can be modeled under the right conditions.
    \item Systematic errors in EM-counterpart models can significantly bias cosmological measurements.
    \item The degeneracy between the inferred underlying NS merger population and the cosmological parameters.
    \item EM-counterpart models which have been trained using a specific cosmology cannot be used since the cosmology itself is being inferred.
\end{itemize}
In all, the thunder-and-lightning method potentially has an exciting role to play in the ongoing cosmic controversy, but there are significant modeling challenges that must be overcome first.

\acknowledgments
I would like to thank the LVC for publicly releasing posterior samples and the ENGRAVE Collaboration for curating the X-Shooter AT2017gfo spectra. I acknowledge Michael Coughlin and Antonella Palmese for useful conversations and for reading a draft of this paper.


\software{Matplotlib \citep{Hunter:2007}, Numpy \citep{Numpy}, Scipy \citep{Scipy}, gwemlightcurves (https://gwemlightcurves.github.io/)}


\bibliography{sample63}{}
\bibliographystyle{aasjournal}

\end{document}